# Downsizing Machine Learning Models by Optimization through Ising Models

Author: Yasuharu Okamoto[1,2]

[1]Secure System Platform Research Laboratories, NEC Corporation, 1753 Shimonumabe, Nakahara-ku, Kawasaki, Kanagawa 211-8666, Japan [2]NEC-AIST Quantum Technology Cooperative Research Laboratories, 1-1-1 Umezono, Tsukuba, Ibaraki, 305-8568, Japan

Corresponding author: E-mail: y-okamoto@aist.go.jp

## Abstract

This paper introduces a technique to enhance the efficiency of quadratic machine learning models, particularly Field-Aware Factorization Machines (FFMs) handling binary data. Our approach strategically reduces model size through optimized feature selection based on the Ising model, maintaining comparable accuracy to the original model. By exploiting the adjustability of FFM's cross-term weights during a novel two-stage training process, we demonstrate a significant improvement in overall computational efficiency.

## Introduction

 Hybrid approaches for solving combinatorial optimization problems by integrating multiple methods have been attracting attention in the field of quantum computation. When using computational platforms such as quantum computers/quantum annealers, we resort to a hybrid quantum-classical approach for dealing with real-world problems because the size of the problem to which quantum computation can be applied is currently quite small. Even a trapped-ion quantum computer with relatively high fidelity can handle at most about 50 bits, and the number of gate stages is limited to a few dozen. On the other hand, although D-Wave's quantum annealing machine has more than 5,000 qubits [1], the number of qubits is far less when the all-to-all connection model needs to be handled. A hybrid quantum-classical approach is therefore indispensable.

In a recent study, a $k$-regular graph with $n$ vertices can be decomposed into a graph with at most $nk/((k+1))$ vertices, and an optimal solution to the $n$ = 100 Max-cut problem was found by combining use of Gurobi (a mixed integer programming solver) and a trapped-ion quantum computer [2]. D-Wave has also focused on research in this area, and has previously proposed Qbsolv, which combines quantum annealing and tabu search using classical computers [3] and has recently recommended a group of hybrid solvers [4].

Combinatorial optimization problems tend to increase computation time and decrease solution accuracy as the number of combinations to be searched explodes with the size of the problem. Therefore, if the problem can be divided into several smaller problems, it may be advantageous to solve the smaller problems for both computational resources and accuracy. This is true not only in quantum computation, but also in other fields, such as divide-and-conquer methods and dynamic programing on classical computers.

However, there is no universal solution in the field of optimization, as known as the "no-free-lunch theorem" [5]. For example, geometric problem partitioning often assumes a specific or special graph structure or problem setting, such as a regular graph, as mentioned above.

Field-aware factorization machines (FFMs) [6] and its ancestor FM [7] are often used as black-box optimization methods based on the Ising model. This paper describes a method for extracting important elements from the weight matrix of FFM and reconstruct them into small matrices. Specifically, assume that the original FFM model is represented by a matrix $Q$ of size $(N \times N)$, and consider decomposing it into a sum of $m$ small matrices of size $(M \times M)$. If the components of the small matrices have no common parts, the computation of the small matrices can be performed independently (or parallelized if computing resources are available), leading to a reduction in the elapsed time. Besides, if the prediction model of FFM can be made smaller, the prediction processing with terminals is also expected.

This paper describes a method for decomposing the original FFM training model into smaller models while minimizing the loss of accuracy. Specifically, the method consists of repeating FFM learning and feature selection twice. This enables the construction of a model with almost the same accuracy using only 26.2% of the full matrix elements of original large matrix $Q$ concerning the example studied in this paper.

# Computational methods

**Data and Preprocessing:** We used data from 442 diabetic patients, refining it for analysis with scikit-learn [8]. The dataset included explanatory variables like Body Mass Index (BMI) and cholesterol levels, alongside the degree of disease progression over one year as the response variable. To prepare for the Field-Aware Factorization Machine (FFM) model adhering to the Ising framework, quantitative variables like BMI and cholesterol were discretized into binary features based on quartile bins. Originally, the dataset comprised ten variables, nine of which were quantitative. These variables were binned into four quartiles, resulting in a transformed total of ($N = 38$) features. We maintained a training-to-test data split ratio of 3:1, yielding 331 training samples. The stochastic gradient descent (SGD) method was employed to ascertain the latent vectors and to compute the second-order coefficients inherent to the FFM from this training set.

**FFM Model:** Typically, an FFM includes constant, first-order, and second-order terms. In our approach, we focused on constructing an FFM model that prioritizes the handling of second-order terms by incorporating only them in combination with the constant term, while excluding the first-order terms. Thus, the predicted outcome $\hat{y}_k$ for the $k$-th data point is described as:

$$\hat{y}_k = w^{(0)} + \sum_{l_2 > l_1}^{D} w^{(2)}_{l_1 l_2} q_{k l_1} q_{k l_2} \qquad (1)$$

We applied this model as its efficiency remained stable despite omitting the linear elements (see Appendix figure A1 for further insights). The binary variable $q_{kl}$ indicates the presence (1) or absence (0) of a feature. The parameter weights $w^{(0)}$ and $w^{(2)}_{l_1 l_2}$ are defined through FFM training. To approximate these weights, the model leverages the inner product of latent vectors:

$$w^{(2)}_{l_1 l_2} \sim \sum_{m=1}^{K} v^{m}_{l_1 f(l_2)} v^{m}_{l_2 f(l_1)} \qquad (2)$$

Here, $f(l)$ correlates each feature with its respective field. Optimization of $w^{(0)}$ and $v^{m}_{lf}$ parameters is achieved by employing SGD to minimize the residual sum of squares depicted in Equation 1. Unless specified, the model converges efficiently at 300 epochs.

**Optimization Through Ising Model:** This method decomposes features into $m$ subsets $S_g$ ($g = 1, \cdots, m$), each containing $M$ features chosen from the full $N$ features. Focusing on potential cross terms within subsets, each subset is limited to $_M C_2 = M(M-1)/2$ cross terms. We enforced constraints ensuring each feature is assigned to only one subset. The predicted value based on the decomposition is:

$$\hat{y}_k^{S'} = w^{(0)} + \sum_{l_2 > l_1}^{D} w^{(2)}_{l_1 l_2} q_{k l_1} q_{k l_2} \chi_{S'}([l_1, l_2]) \qquad (3)$$

A critical task is selecting cross terms that maintain meaningful reduction in computational requirement while preserving model integrity. The optimization objective maximizes the summed absolute weights $w^{(2)}_{l_1 l_2}$, as formulated below(the problem is treated as a minimization

problem with a negative sign instead of maximization problem):

$$\mathcal{T} = -\sum_g \left(\sum_{i,j} \left|w_{i,j}^{(2)}\right| x_{ig} x_{jg}\right) + A \sum_i \left(\sum_g x_{ig}\right)\left(\sum_{g'} x_{ig'} - 1\right) + B \sum_g \left(\sum_i x_{ig} - M\right)^2 \quad (5)$$

The variable $x_{ig}$ is binary, equaling 1 if the *i*-th feature belongs to subset $S_g$, and 0 otherwise. This variable is subject to optimization within the Ising model framework. Its relationship with the indicator function $\chi_{S'}([i,j])$, as previously used in eq. (3), is:

$$\chi_{S'}([i,j]) = \sum_g x_{ig} x_{jg} \quad (6)$$

The first term on the right side of Equation (5) serves to maximize the sum of the absolute values of the weights $w_{i,j}^{(2)}$. The second term enforces the constraint that each feature belongs to at most one subset. The third term constrains the number of elements within each subset to be *M*. The optimization challenge lies in selecting an appropriate value for *m*, representing the number of subsets. In practice, this optimization problem is tackled through an iterative approach, exploring various combinations of (*M*, *m*). A simultaneous evaluation of all combinations is generally infeasible due to the rapidly escalating number of variables, thus hindering a realistic computational strategy.

## Results and discussion

First, to examine whether it is possible to decompose a whole FFM model into a sum of subsets reproducing the original model, we examined the effect of the number of subsets *m* and their number of elements *M* on the model's accuracy. The loss function is defined as the sum of the residual squares of the true and predicted values. Figure 1A shows the loss behavior when *M* is varied for *m* = 1 - 5. The dashed horizontal line represents the value when all (nonzero) matrix elements are considered, *i.e.*, the value from the original model.

The maximum possible value of *M* for each *m* is $M_{max}(m) = floor(N/m)$. Naturally, the loss decreases as *M* increases, and for the same *M*, the loss decreases for larger *m*. In this case, the larger *m* is, the more matrix elements are covered. However, since *M* has an upper bound, and $M_{max}(m)$ decreases as *m* increases, the number of matrix elements in the cross term generated from the subset decreases as *m* increases. For example, when (*M*, *m*)=(7, 5), only about 15% of the original matrix elements are considered, so the loss cannot be reduced sufficiently. We found that $m = floor(N/M_{max})$ a good choice.

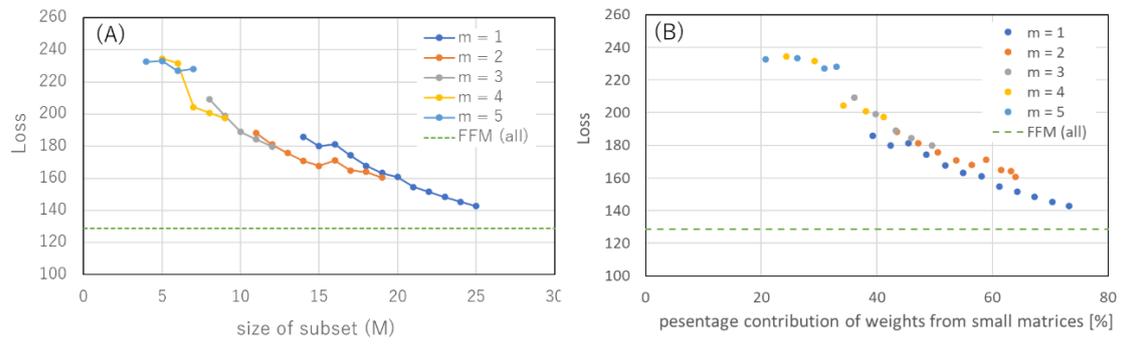

Figure 1: (A) Relationship between the number of subsets $m$ and the number of their elements $M$ and loss. (B) Relationship between loss and the ratio of the absolute value of the selected weights to all elements.

Figure 1B shows the ratio of the sum of the absolute weights of the cross terms generated from the subsets to the sum of the absolute weights of all cross terms in the original FFM model (horizontal axis) and the relationship between loss (vertical axis). As mentioned above, in this calculation, the optimal subset combination is selected by maximizing the sum of absolute values of the weights for a given $(M, m)$. Theoretically, it is desirable to select a combination of matrix elements that minimizes the sum of squared residuals itself. In this case, the objective function contains a fourth-order term since the quadratic equation is squared, which we can solve by introducing auxiliary variables. However, the increase in the number of binary variables to be optimized increases the computation time and decreases the simplicity of the computation. Figure 1B shows that when the sum of the contributions from the subset exceeds 30% of the total, the loss, defined as the sum of the squared residuals of the true and predicted values, decreases almost linearly with the ratio of these weights, indicating that the sum of the absolute values of the cross terms generated from the subset is a good surrogate for the sum of the squared residuals.

Figure 1A shows that the loss decreases when a subset contains many elements. However, we assume that there are situations where a set with a sufficient number of elements cannot be included due to some (hardware) restriction in quantum annealers, the following discussion focuses on the case $(M, m)=(14, 2)$.

Figures 2A and 2B display the scatter plots of the true and predicted values, i.e., $(y_k, \hat{y}_k^{S'})$ and $(y_k, \hat{y}_k^A)$, concerning the training and test data. Here, $\hat{y}_k^{S'}$ is the prediction for the $k$-th data using only the constant term and the cross terms generated from the subset, which are calculated using eq. (3). Conversely, $\hat{y}_k^A$ is the prediction based on the original FFM model computed by eq. (1), when all (nonzero) cross terms are considered before selecting the subset. The red 45° line corresponds to the case where the prediction is perfect. Compared to this, the predicted value $\hat{y}$ tends to be underestimated in the original FFM model in regions where the explained variable $y$ (disease progression) is significant for training and test data. This tendency is also true for models consisting of subsets. The coefficient of determination ($R^2$) of the subset model for the test data (0.398) is slightly lower than that of the original FFM model for all elements (0.428). These results indicate that simply selecting the cross terms is insufficient to reproduce the original model and that some modification is necessary. In this study, the joint matrix of features (the weight matrix of the cross terms) is a parameter determined from the training of FFM, so it is not a problem-dependent quantity like the distance between cities in the travelling salesperson problem (TSP). Thus, we consider them hyperparameters when constructing the Ising model. In the following, we discuss improving the prediction accuracy by re-training the cross terms.

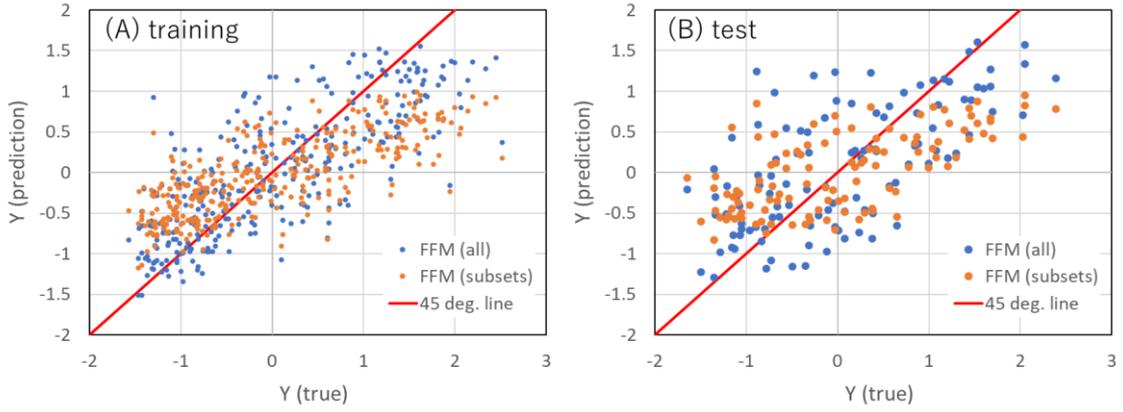

Figure 2: (A) Scatterplot of true values and predictions for the training data. Predictions are based on the original FFM model considering all (nonzero) cross terms (blue circles) and cross terms generated by extracting subsets ($M$, $m$)=(14, 2) from the features of the original FFM model (orange circles). (B) Scatter plots for the test data with the same settings as in (A).

Figure 3 shows a schematic diagram of the re-training procedure. Thus far, we have constructed a learning model by generating a cross-term using a subset of the features obtained in (c) along the process (a) - (c). Here, only the matrix elements ($i$, $j$) corresponding to the selected cross terms are used as training parameters to learn the FFM again, and the improved weights of the cross term (and constant term) are used to reselect the subset by the Ising model (process (d)-(f) in Figure 3).

However, there is a minor problem here. In FFM, the cross terms are not directly treated as training parameters, but the latent vectors are used as training parameters. As shown in eq. (2), the cross-term is computed as the inner product of the latent vectors. Therefore, the learning of latent vectors involves learning the matrix elements ($i$, f(j)) that belong to the same field as $j$, similarly, the matrix elements $(j, \mathrm{f}(i))$ that belong to the same field as $i$ for the cross-term $w_{ij}$. This means that the cross-term elements are not only constructed by the feature selection in the process (c) but also by a few extra matrix elements.

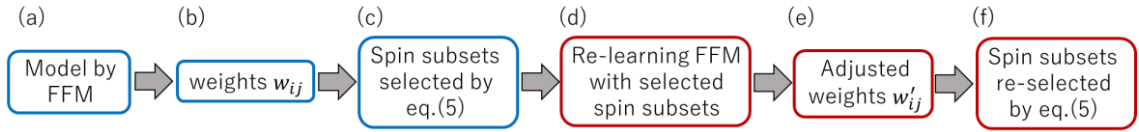

Figure 3: Schematic of selecting a subset of features that reproduce the original FFM model.

Figure 4 shows the relationship between the true values and predictions for the test data for the model constructed by decomposing the weights of the cross terms obtained by such re-training into subsets again. The orange circles are the predicted value from single training session corresponding to process (c) in Fig. 3. On the other hand, the green circles represent the predictions from two training sessions corresponding to process (e). As shown in the figure, the tendency to underestimate the large $y$ region is considerably mitigated for the predictions using two training sessions. The $R^2$ is 0.420, which shows an improvement but is still slightly lower than the original result for all cross terms (0.428).

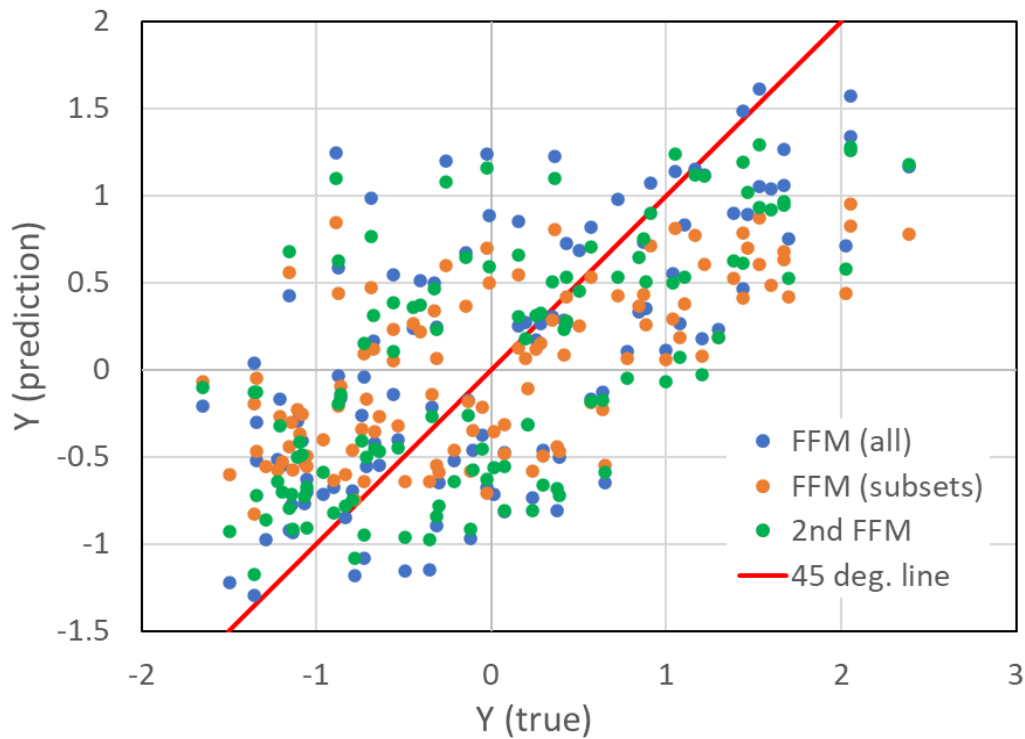

Figure 4: Scatterplot of true values and predictions for the test data. Predictions are from the original FFM model considering all (nonzero) cross terms (blue circles), from cross terms

generated by extracting a subset ($M$, $m$)=(14, 2) from the features of the original FFM model (orange circles), from cross terms generated by re-training the FFM to improve the weights of the cross terms and then extracting a subset ($M$, $m$)=(14, 2) (green circles).

It is noteworthy that the cross term $q_{ki}q_{kj}$ can be directly selected as explanatory variables in the linear regression model [8], the re-training can be performed using only the terms determined by the first feature selection, without including extra terms derived from the same field, as in FFM. The results are shown in Figure 5. In this case, the generalization performance is degraded due to the obvious increase in dispersion with respect to the test data. The $R^2$ is -0.073, which is very poor (it can be negative according to the definition of it in this paper). On the contrary, the $R^2$ for the training data is 0.761, which is considerably larger than the original result for all cross terms (0.619), indicating that a rather severe overfitting has occurred.

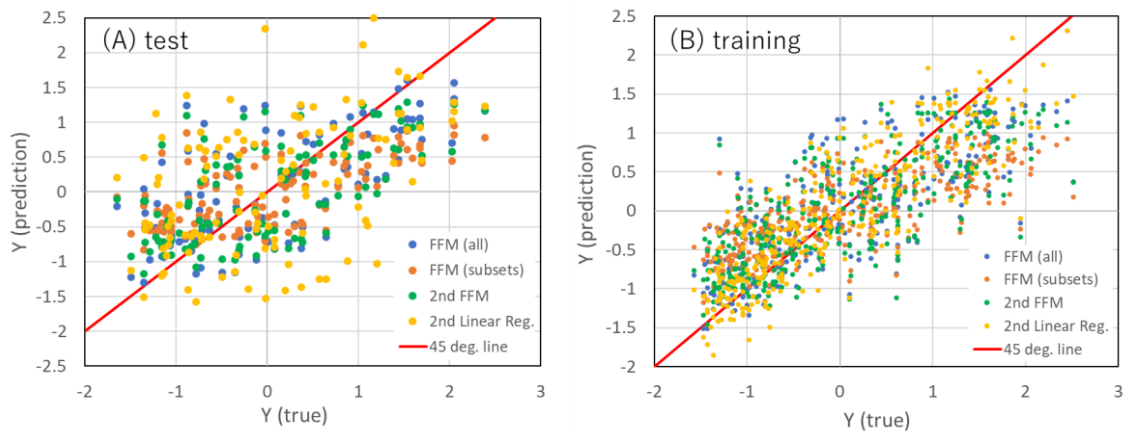

Figure 5: (A) Scatterplot of true values and predictions for test data. (B) Scatter plots of true values and predictions for the training data. The blue, orange, and green circles indicate the same objects as in Figure 4. The yellow circles indicate the second training with linear regression.

Since the result of FFM that has a few extra terms in the second training was good, we consider deliberately increasing the number of features to be trained for the first training session in (c) in Fig. 3. However, we assume that the second feature selection in (e) is limited by hardware and/or other factors, so $M = 14$ is left unchanged. Now, if two subsets are created, the maximum of $M$ is 19 since the total number of features is 38. Table 1 shows the behavior of the $R^2$ for the training and test data by the model obtained from $M = 14$ - 19 for the first training session in (c).

Table 1: Coefficient of determination ($R^2$) when extra elements are included in the first learning session.

| $M$ | 14 | 15 | 16 | 17 | 18 | 19 |
|---|---|---|---|---|---|---|
| $R^2$ (training) | 0.586 | 0.593 | 0.571 | 0.570 | 0.569 | 0.578 |
| $R^2$ (test) | 0.420 | 0.417 | 0.461 | 0.462 | 0.464 | 0.416 |

The $R^2$ for the test data can be regarded as a measure of generalization performance, and it reaches a maximum value of 0.464 when $M = 18$, which is higher than that of the original FFM learning model (0.428) for all (nonzero) cross terms. Figures 6A and 6B show the scatter plots $(y_k, \hat{y}_k^S)$ of the true values and predictions for the training and test data, respectively, when the number of features selected for the first time is $M = 18$. The result shows that the cross terms generated from the subset well reproduce the predictions of the original FFM learning models for all (nonzero) cross terms. It is noteworthy that the first-order approximation curves based on the two distributions for the test data are almost identical. The total number of weight elements in the two upper triangular matrices (excluding the diagonal terms) is 182, compared to 703 elements in the original FFM weight matrix of size $(38 \times 38)$, which means that we have constructed a prediction model with almost the same accuracy.

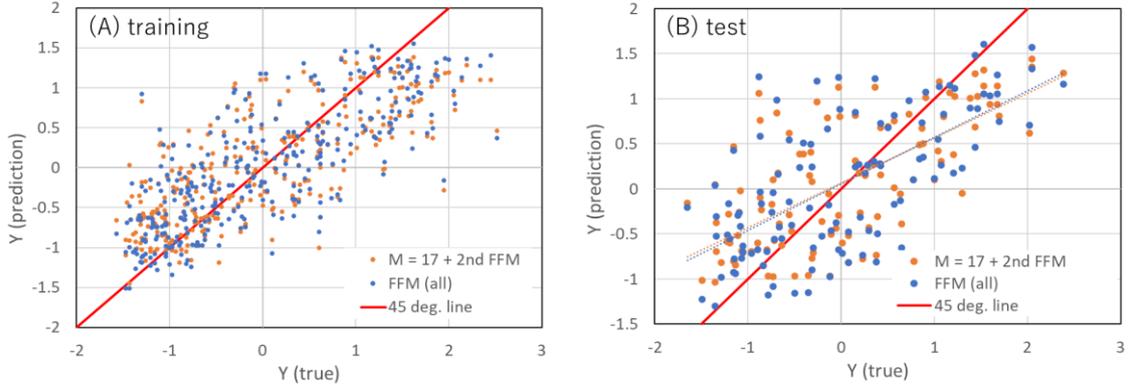

Figure 6: (A) Scatterplot of true values and predictions for the training test data. (B) Scatter plots of the true and predicted values for the test data. The dotted lines are first-order approximations of the distributions of the blue and orange circles, respectively, showing that they almost overlap.

As stated, twice training and feature selection enabled the construction of a small-scale but good-accuracy training model, but the procedure requires two rounds of training, thus it is still not an efficient approach. The first feature selection only extracts effective feature candidates for re-training, so it is not necessary to compute the weights of the cross terms with high precision. Therefore, the first training does not need to converge sufficiently, and the weights of the cross terms are generated from the results at low epochs, and we use them used in the second session. Figure 7 shows the behavior of the loss function in the first FFM training and the $R^2$ of the final result for the test data when the number of epochs of the first training is stopped at 20, 50, 75, and 100 with $M = 18$ (The number of epochs for the second training is 300, and the second selection is the same as before with the condition $(M, m)=(14, 2)$). We found that the result for epoch 100 is consistent with the previous result for epoch 300 in the first training. It can also be seen that the final $R^2$ is better than the original FFM learning models for all cross terms (0.428), even though the loss function for epoch 20 has not converged sufficiently. Thus, it is possible to improve the overall computational efficiency by modifying the learning process.

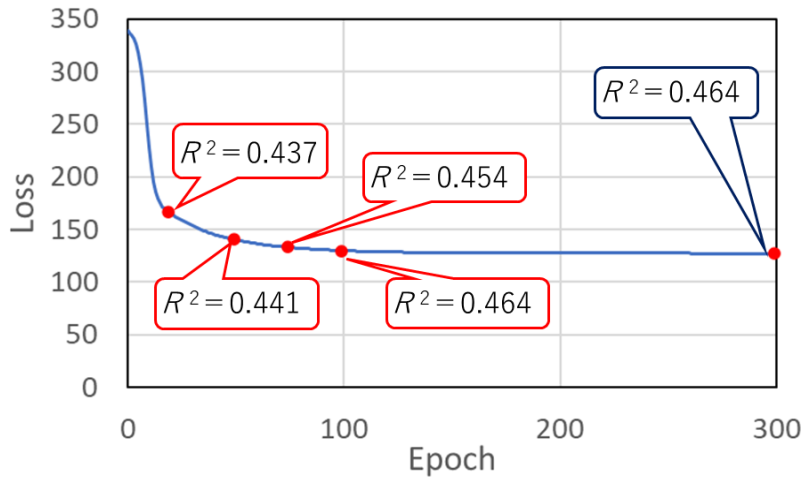

Figure 7: Behavior of loss function in the first FFM training stopped at epochs 20, 50, 75, and 100. Values correspond to final $R^2$ values.

# Summary

For machine learning models with quadratic (cross) terms, such as FFM, when the data are given as 0/1 binary variables, we propose a method that can significantly reduce the model size by selecting appropriate features through optimization based on the Ising model, without deteriorating the generalization performance compared to the original model. In optimization problems such as FFM, the weights of the cross terms of the models are quantities corresponding to the distance between cities in the TSP. However, in TSP, this distance is given by the problem and cannot be changed, while in FFM, the cross terms are learning parameters and can be adjusted. Therefore, by training the FFM twice and selecting features, we were able to construct a training model that is smaller in size while maintaining accuracy. Furthermore, we showed that it is possible to improve the overall computational efficiency by modifying the learning process.

## Acknowledgements

This paper was (partly) based on the results obtained from a project, JPNP16007, commissioned by the New Energy and Industrial Technology Development Organization (NEDO), Japan.

# Appendix figure A1

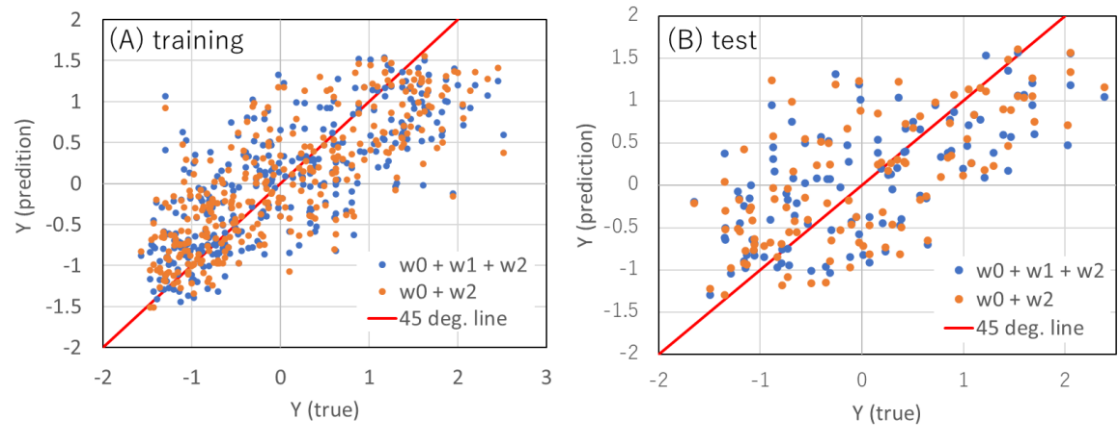

Figure A1: (A) Scatterplot of true values and predictions for the training data. (B) Scatterplot of true values and predictions for the test data. The blue circles are the usual FFM model with constant, linear, and quadratic terms. The orange circles are the FFM model with constant terms and quadratic terms only, which is the focus of this paper.